\documentclass[10pt,preprint]{aastex}
\usepackage{amssymb,graphicx,longtable,lscape,natbib,multirow}
\usepackage{wasysym}
\usepackage{color}

\def\msol{\hbox{$\rm\thinspace M_{\odot}\thinspace$}} 
\def\sol{\hbox{$\rm\thinspace _{\odot}\thinspace$}} 
\def\etal{{\it et al.\thinspace}}
\def\eg{{\it e.g.\ }}

\newcommand{\be}{\begin{equation}}
\newcommand{\ba}{\begin{eqnarray}}
\newcommand{\ee}{\end{equation}}
\newcommand{\ea}{\end{eqnarray}}

\newcommand{\nuclei}[2]{\ensuremath{\mathrm{^{#1}#2}}}

\newcommand{\helium}[1][4]{\nuclei{#1}{He}}

\newcommand{\carbon}[1][12]{\nuclei{#1}{C}}

\newcommand{\oxygen}[1][16]{\nuclei{#1}{O}}

\newcommand{\nickel}[1][58]{\nuclei{#1}{Ni}}

\begin{document}

\title{Remnants of Binary White Dwarf Mergers}
\author{Cody Raskin\altaffilmark{1}, Evan Scannapieco\altaffilmark{1}, Chris Fryer\altaffilmark{2}, Gabriel Rockefeller\altaffilmark{2}, \& F.X. Timmes\altaffilmark{1,3}}
\altaffiltext{1}{School of Earth and Space Exploration,  Arizona State University, P.O.  Box 871404, Tempe, AZ, 85287-1404} 
\altaffiltext{2}{Los Alamos National Laboratories, Los Alamos, NM 87545}
\altaffiltext{3}{The Joint Institute for Nuclear Astrophysics}

\begin{abstract}

We carry out a comprehensive smooth particle hydrodynamics simulation survey of double-degenerate white dwarf binary mergers of varying mass combinations in order to establish correspondence between initial conditions and remnant configurations. We find that all but one of our simulation remnants share general properties such as a cold, degenerate core surrounded by a hot disk, while our least massive pair of stars forms only a hot disk. We characterize our remnant configurations by the core mass, the rotational velocity of the core, and the half-mass radius of the disk. We also find that some of our simulations with very massive constituent stars exhibit helium detonations on the surface of the primary star before complete disruption of the secondary. However, these helium detonations are insufficiently energetic to ignite carbon, and so do not lead to prompt carbon detonations.

\end{abstract}

\section{Introduction}

Type Ia supernovae are commonly accepted to be the observed transient produced after a thermonuclear detonation inside a white dwarf star. While the preferred mechanism for producing SNeIa involves accretion from an evolved main sequence star onto a white dwarf (Whelan \& Iben 1973; Nomoto 1982; Hillebrandt \& Niemeyer 2000), the observed SNeIa rate is incompatible with the narrow range of helium accretion rates that initiate a carbon detonation as opposed to accretion induced collapse or classical novae (Nomoto \& Kondo 1991; Hardin \etal 2000; Pain \etal 2002; Ruiter \etal 2009). Moreover, many recent observations of abnormally luminous SNeIa have been interpreted as having derived from double-degenerate systems involving two white dwarfs. 

For example, photometric observations of SN~2007if suggest 1.6$\pm$0.1\msol of \nickel[56] was formed, implying a progenitor mass of 2.4$\pm$0.2\msol (Scalzo \etal 2010), which is well above the Chandrasekhar limit (Chandrasekhar 1931). Spectroscopic observations of SN~2009dc suggest $\apprge1.2$\msol of \nickel[56] (Tanaka \etal 2010), depending on the assumed dust absorption. Since 0.92\msol of \nickel[56] is the greatest yield a Chandrasekhar mass can produce (Khokhlov \etal 1993), this yield also implies a super-Chandrasekhar progenitor mass. And observations of SN~2003fg by Howell \etal (2006) and of SN~2006gz by Hicken \etal (2007) infer $\sim1.3$\msol of \nickel[56] each.

Generally, for the purposes of cosmological measurements, obvious double-degenerate candidates are excluded from SNeIa surveys. The Phillips relation, or the width-luminosity relation (WLR), which established SNeIa as standard candles (Phillips 1993), relates the peak luminosity of a SNIa to the change in magnitude after 15 days. The WLR for standard SNeIa indicates that SNeIa with bright peak magnitudes also decay at a slower rate than dimmer SNeIa. This is often thought to be the result of a relationship between the \nickel[56] yield and the opacity of the ejecta material, assuming a total mass not exceeding the Chandrasekhar mass. However, a double-degenerate system may have up to two times the Chandrasekhar mass, and so the relationship between the \nickel[56] yield and the ejecta opacity need not be similar to single-degenerate scenarios, and the WLR may not be applicable to these SNeIa. In fact, it is more likely that for a given \nickel[56] production and energy deposition history, an increased ejecta mass results in an increased opacity, reducing the peak magnitude and broadening the lightcurve in a fashion that is the inverse of the standard WLR (Pinto \& Eastman 2000; Mazzali \etal 2001; Mazzali \& Podsiadlowski 2006; Kasen, R\"opke, \& Woosley 2009). 

Complicating matters is the possibility that SNeIa deriving from double-degenerate progenitors can have ordinary \nickel[56] yields, depending on the progenitor mechanism (see \eg collisional mechanisms in Raskin \etal 2009, 2010 and Rosswog \etal 2009) and the final, central densities of the degenerate material before ignition. Thus, double-degenerate SNeIa may be masquerading as typical SNeIa. If they do not conform to the WLR, they may introduce systematic errors into cosmological surveys. In order to reduce the scatter in the Hubble diagram, we must first establish whether double-degenerate SNeIa are standardizable, and if not, we must identify the tell-tale signatures of a double-degenerate progenitor mechanism.
 


The most probable double-degenerate progenitor scenario involves two white dwarfs in a tight binary, though other progenitor systems have been considered (Benz \etal 1989a; Raskin \etal 2009; Rosswog \etal 2009). Binary white dwarf systems were first seriously explored as plausible SNeIa progenitors by Iben \& Tutukov (1984) and Webbink (1984). In such a system, tidal dissipation and gravitational radiation drive the binary pair into an ever closer orbit. Eventually, the least massive white dwarf, being physically larger as $R\propto M^{-1/3}$, overflows its Roche lobe and begins to accrete material onto the primary, or more massive companion star. For many mass combinations, this is a fundamentally unstable process in which the loss of mass from the secondary causes it to outgrow its Roche lobe faster than its orbit widens due to conservation of angular momentum (see \eg Marsh \etal 2004). 

Benz \etal (1990) performed one of the first simulations of double-degenerate mergers, examining a binary system consisting of a 1.2\msol white dwarf primary and a 0.9\msol white dwarf secondary. They used a smooth particle hydrodynamics code with 3000 particles per star and found that the merger remnant consisted of a 1.7\msol core surrounded by a rotationally supported disk. This is more massive than the Chandrasekhar limit, but they concluded that the central object was not entirely degenerate, having been considerably heated, and thus, much of the object's support against gravitational collapse came simply from thermal pressure. 

Since this pioneering work, others have revisited the white dwarf merger scenario with up-to-date simulation codes and higher resolutions than were possible in the past (see \eg Yoon \etal 2007; Lor\'{e}n-Aguilar \etal 2009). In this paper, we revisit white dwarf mergers simulations, examining a wide range of possible mass combinations with high resolution, more accurate initial conditions, and up-to-date physics for the equation of state and nuclear reaction network. The structure of this paper is as follows. In \S2, we outline our methods and initial conditions. We discuss the results of our simulations in \S3, and compare our work to previous studies of white dwarf mergers in \S4. Finally, in \S5, we summarize our results and conclusions and discuss possible avenues for advancing remnant evolution in future studies.

\section{Method \& Initial Conditions}

As in our previous work on white dwarf collisions (Raskin \etal 2010), we employ the 3D smooth particle hydrodynamics code SNSPH (Fryer \etal 2006). Particle codes are ideally suited to simulating binary systems as they excel at conserving angular momentum, which is crucial for properly simulating accretion flows in binary systems. We include a Helmholtz free energy equation of state (EOS; Timmes \& Arnett 1999; Timmes \& Swesty 2000) that spans a range of possible thermodynamic conditions, from cold, electron-degenerate gas, to hot, non-degenerate gas, and includes coulomb corrections and photon pressures. For nucleosynthesis, we use a 13-isotope, $\alpha$-chain nuclear network with a hybrid photo-disintegration capture scheme as described in Raskin \etal (2010), which utilizes a similar method for sub-cycling the reaction rates to what is implemented in MAESTRO (Nonaka \etal 2010). We limit our hydrodynamic time step to that which results in at most a 30\% change in internal energy ($u$) due to nuclear reactions as this was found in Raskin \etal (2010) to strike the right balance between accuracy of the result and efficiency of the calculation. If nuclear reactions are not important, the time step is not allowed to grow larger than what is derived from the Courant-Friedrichs-Lewey condition. 

For our constant mass particle arrangements, we use Weighted Voronoi Tessellations (WVT; Diehl \& Statler 2006), which ensures the lowest energy particle configuration for our stars. We initiate each of our carbon-oxygen white dwarfs with $5\times10^5$ particles per star and a uniform composition of 50\% \carbon[12] and 50\% \oxygen[16] plus a thin atmosphere (no greater than $q($He$)=M_{\rm He}/M<0.025$) of pure \helium[4] that approximates atmospheric mass estimates from stellar evolution models (Iben \& Tutukov 1985; Salaris \etal 2000). The \helium[4] masses for each star are given in Table \ref{table:helium}. Each of our stars is relaxed to an isothermal temperature of roughly $10^7$ K. The noise in the temperatures of the isothermal stars due to the particle nature of our simulations is of order $\sim10^6$ K. 

\begin{table}[ht]
\caption{Atmospheric helium masses for each of the stars used in our simulations. All masses are in solar mass units.}
\centering
\begin{tabular}{c | c}
\hline\hline
$m_{\rm tot}$ & $m_{\rm He}$\\
\hline
0.64 & 0.015\\
0.81 & 0.013\\
0.96 & 0.012\\
1.06 & 0.013\\
\hline
\end{tabular}
\label{table:helium}
\end{table}

There is some debate over whether white dwarfs in tight binaries will be tidally locked. The time scales for inspiral and for spinning up the white dwarfs are roughly commensurate. Yoon \etal (2007) initialized their binary systems with tidally locked white dwarfs, while Lor\'{e}n-Aguilar \etal (2009) assumed they  were not tidally locked. Since many close white dwarf binaries are expected to be tidally locked (Claret \& Cunha 1997; Marsh \etal 2004; Geier \etal 2008), we make a conscious choice to simulate our binaries in a tidally locked configuration. 

We initially allow our stars to fall from rest toward each other until there is noticeable tidal distortion in the least massive star. This is done via an iterative process whereby the stars free-fall, and at intervals of 5s each, all particles have their velocities reset to zero and free-fall begins again. This ensures that the binary system is in a meta-stable configuration and that the stars are still in hydrostatic equilibrium after tidal distortions have first appeared. 

\begin{figure}[ht]
\centering
\includegraphics[width=0.4\textwidth]{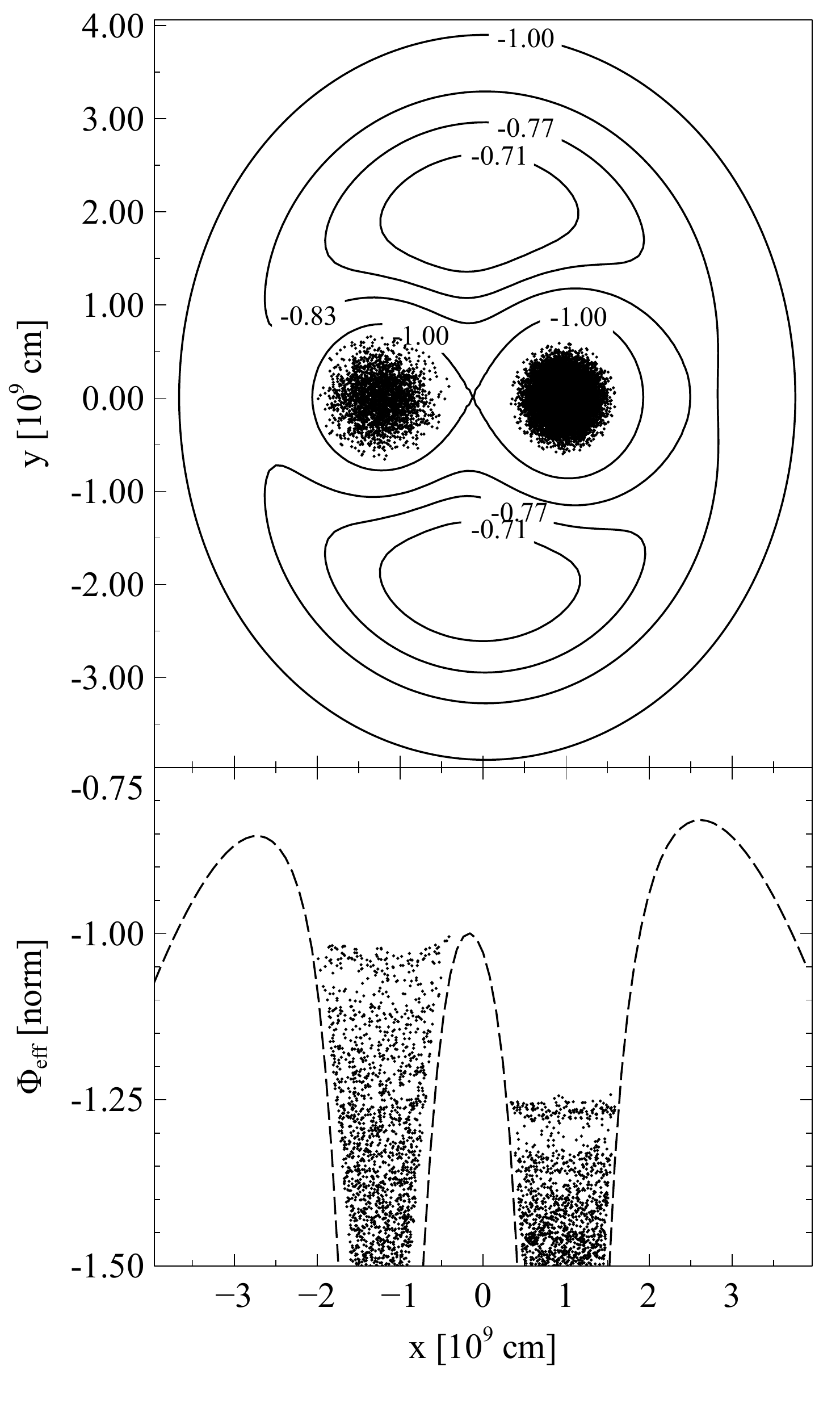}
\caption{{Top panel:} Particle positions for a representative subsample of all particles, sampled uniformly from the entire distribution, overlaid on top of a contour map of the rotating frame effective potential in the initial conditions ($t=0$) of the 0.64\msol + 0.81\msol merger simulation. The effective potential is normalized to $-1$ at the Lagrange point between the two stars (L1). {Bottom panel:} Particle $x$-positions vs. effective potentials are plotted with the derived potential for the same conditions and normalization as the top panel. {Inset:} Same as bottom panel, but rescaled to better demonstrate the potential difference between the particles of the 0.64\msol star and the point of Roche lobe contact.}
\label{fig:lagrange}
\end{figure}

This method avoids stellar oscillations and unnaturally high accretion rates that result from the secondary white dwarf overflowing its Roche lobe immediately upon the start of the simulation. Figure \ref{fig:lagrange} shows the effective potential at $t=0$ in the rotating frame of the 0.64\msol + 0.81\msol merger simulation, where $\Phi_{eff} = \Phi_g - 1/2\Omega^2r^2$ normalized to $1.79\times10^{17}$ erg/g with $\Omega$ being the orbital velocity. As is evident, none of the particles in the 0.64\msol star lie outside of the Roche lobe, and no particle has a higher potential than the potential wall between the two stars. Dan \etal (2011) discuss how previous efforts in this area have overestimated the accretion rates due to inappropriate initial conditions, and this can have implications for the final arrangement of the system and on the long-term evolution prospects.

Our iterative method for creating the initial conditions of the mergers avoids immediate and rapid mass-transfer while also placing the stars very near the stage where mass transfer would naturally begin. At this stage, the stars are set into their tidally locked orbits about their common center of mass with the initial periods given in Table \ref{table:grid} for each of our chosen mass combinations. Over several orbits, the tidal distortions grow and material from the secondary star overflows its Roche lobe and mass transfer begins. Our mass combinations all lie safely in the regime of unstable mass transfer (Marsh \etal 2004), and so resolving a steady accretion stream that could possibly lead to stable accretion using a discretized method like SPH is not important.

\begin{table}[ht]
\caption{{Simulated binary mass pairs and their initial orbital periods ($\tau_0$). All masses are solar.}}
\centering
\begin{tabular}{c | c c | c c | c}
\hline\hline
\# & $m_1$ & $m_2$ & $m_{tot}$ & $m_1/m_2$ &  $\tau_0$ [s]\\
\hline
1& 	0.64 & 	0.64 & 	1.28 & 	1.00 & 	47.61\\
2& 	0.64 & 	0.81 & 	1.45 & 	0.79 &	46.99\\
3& 	0.64 & 	0.96 & 	1.60 & 	0.67 & 	42.71\\
4& 	0.64 & 	1.06 & 	1.70 & 	0.60 & 	43.99\\
5& 	0.81 & 	0.81 & 	1.62 & 	1.00 &	28.72\\
6& 	0.81 & 	0.96 & 	1.77 & 	0.84 & 	27.83\\
7& 	0.81 & 	1.06 & 	1.87 & 	0.76 & 	25.34\\
8& 	0.96 & 	0.96 & 	1.92 & 	1.00 & 	21.50\\
9& 	0.96 & 	1.06 & 	2.02 & 	0.90 & 	17.95\\
10&	1.06 &	1.06 &	2.12 &	1.00 &	17.56\\
\hline
\end{tabular}
\label{table:grid}
\end{table}


Pakmor \etal (2010) also explored a double degenerate merger of two 0.9\msol white dwarfs in a hybrid SPH/grid-code simulation study that was the first of its kind to track a merger from the initial inspiral to the final detonation and subsequent homologous expansion phase. They found that if such a merger features a prompt detonation before settling into a meta-stable core-disk configuration, the resultant supernova would be exceptionally dim, appearing much like 1991bg. However, the detonation in their simulations was inserted by hand at the location at which the large-scale conditions matched those found in offline calculations to produce microphysical conditions necessary for detonation. 

At first glance, this is a reasonable approach since initiations of detonations are often difficult to resolve in large hydrodynamical simulations. While the conditions across a single SPH particle may not be sufficient to initiate a detonation on their own, it is reasonable to suspect that temperature fluctuations that are unresolved in SPH could lead to a detonation that would not be captured in a simulation like this. However, Seitenzahl \etal (2009) have demonstrated that spontaneous detonations of \carbon[12]/\oxygen[16] material depend very sensitively on both the local thermodynamic conditions and the conditions of the surrounding medium. They find that unless the surrounding medium is also $>10^9$K, thermodynamic conditions that would otherwise lead to a detonation (\eg $2.9\times10^9$ K and 3.8$\times10^6$ g cm$^{-3}$ as was used in Pakmor \etal (2010)) would require pre-detonation hotspots to be of order $10^7$ cm in extent. In fact, many of our simulations produced similar detonation preconditions in local hotspots at the surface of the accretor star as in Pakmor \etal (2010), but these hotspots were isolated in a medium of relatively colder material. Moreover, they were not gravitationally confined as they were part of an orbiting flow. Instead, they merely expanded outward hydrodynamically and floated away from the surface, cooling rapidly. This rapid cooling quenched any significant carbon burning and staved off spontaneous carbon detonations arising naturally in the simulations. More importantly, the initial conditions used in Pakmor \etal (2010) vastly overestimate the accretion flow, and thus the accretion shock and shock heating of the disk material. For these reasons, we choose not to insert detonations into those hotspots.

\section{Results \& Analysis}

\begin{figure*}[ht]
\centering
\includegraphics[width=0.85\textwidth]{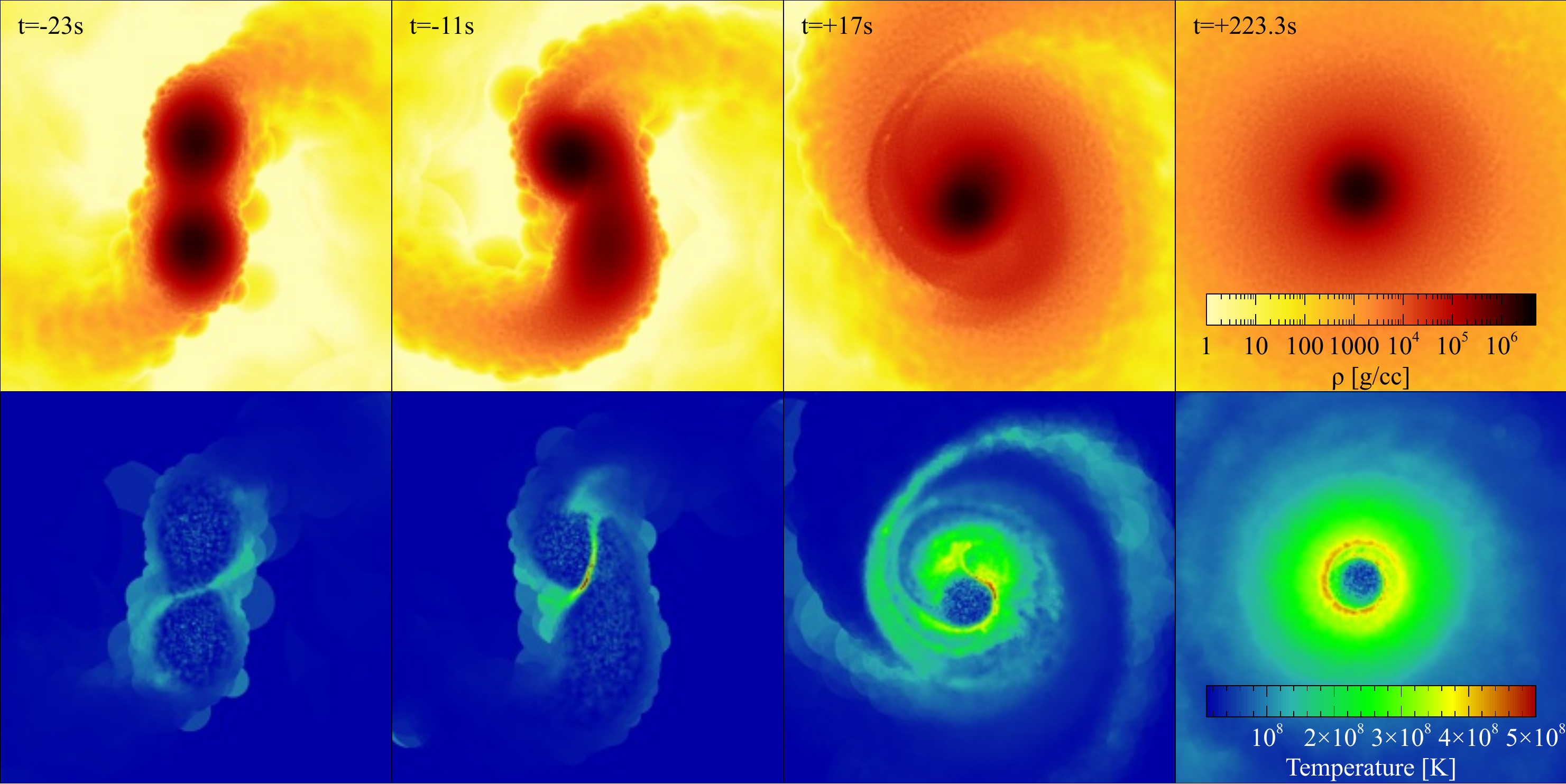}
\caption{Snapshots in time of simulation 5, 0.8\msol$\times2$, with the time coordinate centered on the moment of complete destruction of the secondary star. The top four images are density maps of slices in the $x$-$y$ plane, while the bottom four images are temperature maps of the same slice.}
\label{fig:0p8x2}
\end{figure*}

In our simulations of equal-mass binaries, such as simulation 5 with 0.8\msol$\times2$, depicted in Figure \ref{fig:0p8x2}, there is first a protracted accretion phase that persists for several orbits. During this time, both stars lose material, and thus angular momentum, through the L2 and L3 points. This brings the stars closer together, hastening their mass loss. The helium atmospheres that the stars were initially given also burns fairly rapidly to carbon and oxygen during this phase. Finally, one of the stars is disrupted entirely, owing primarily to slight inhomogeneities in the constituent stars. At this stage, $\approx$2\% of the carbon is burned into heavier, silicon-group elements, but the iron-group yield from the merger is negligible. 

\begin{figure*}[ht]
\centering
\includegraphics[width=0.85\textwidth]{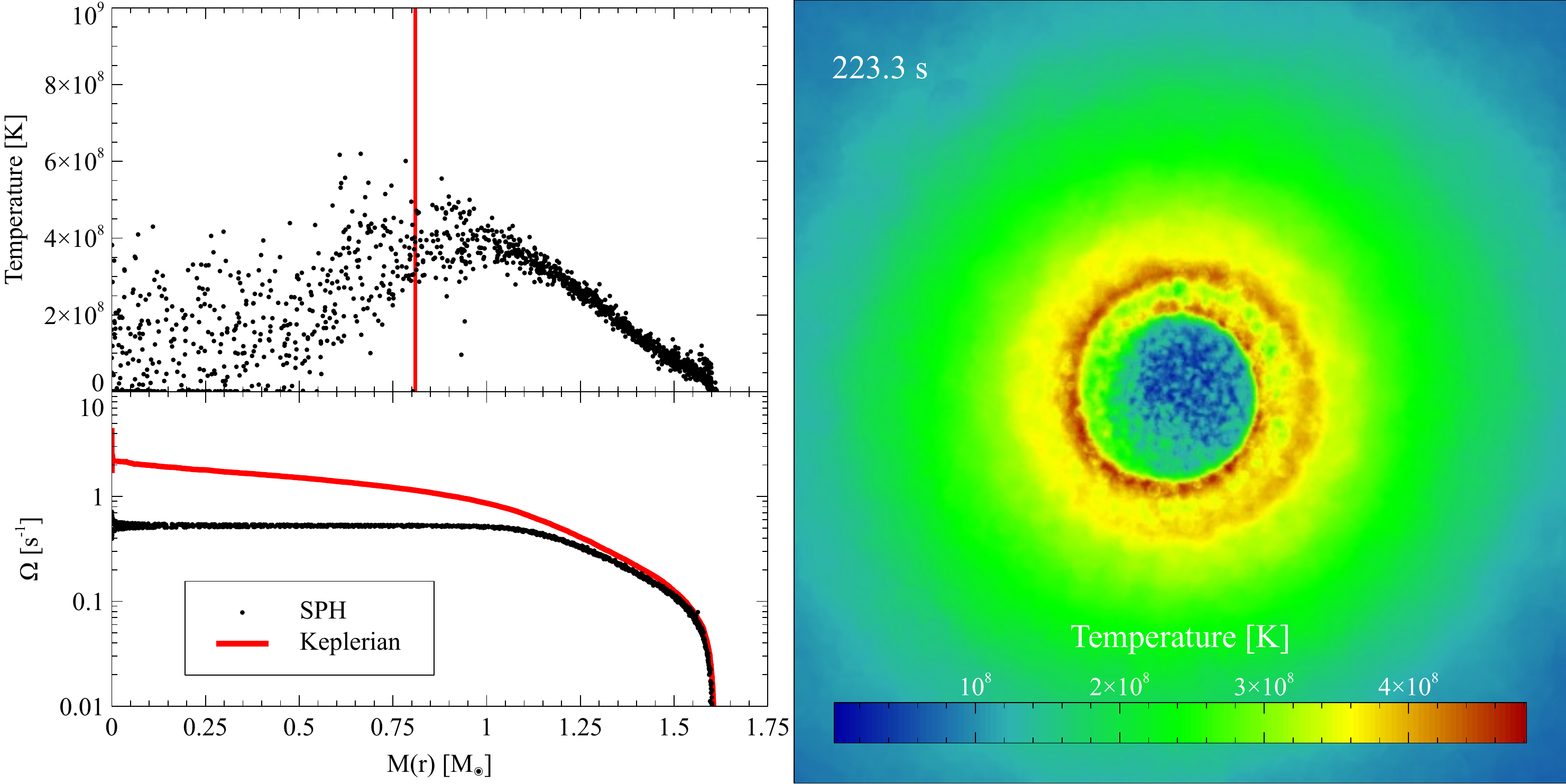}
\caption{Left Panel: Temperature and angular velocity as functions of mass coordinate for the remnant configuration of simulation 5, 0.8\msol$\times2$ at 223.3s after complete destruction of the secondary. The red vertical line indicates the progenitor mass of the primary star. Right Panel: A slice in the $x$-$y$ plane of the temperature of the remnant.}
\label{fig:5rem}
\end{figure*}

After many orbital periods, the system settles into a meta-stable configuration with a dense core, a shock-heated, sub-keplerian disk, and a semi-degenerate interface between them. As Figure \ref{fig:5rem} indicates, much of the core has been heated slightly to $\approx$2$\times10^8$ K, while the interface reaches temperatures near $10^9$ K. The core and interface also exhibit solid-body rotation out to $\approx$1.05\msol. 

\begin{figure}[ht]
\centering
\includegraphics[width=0.45\textwidth]{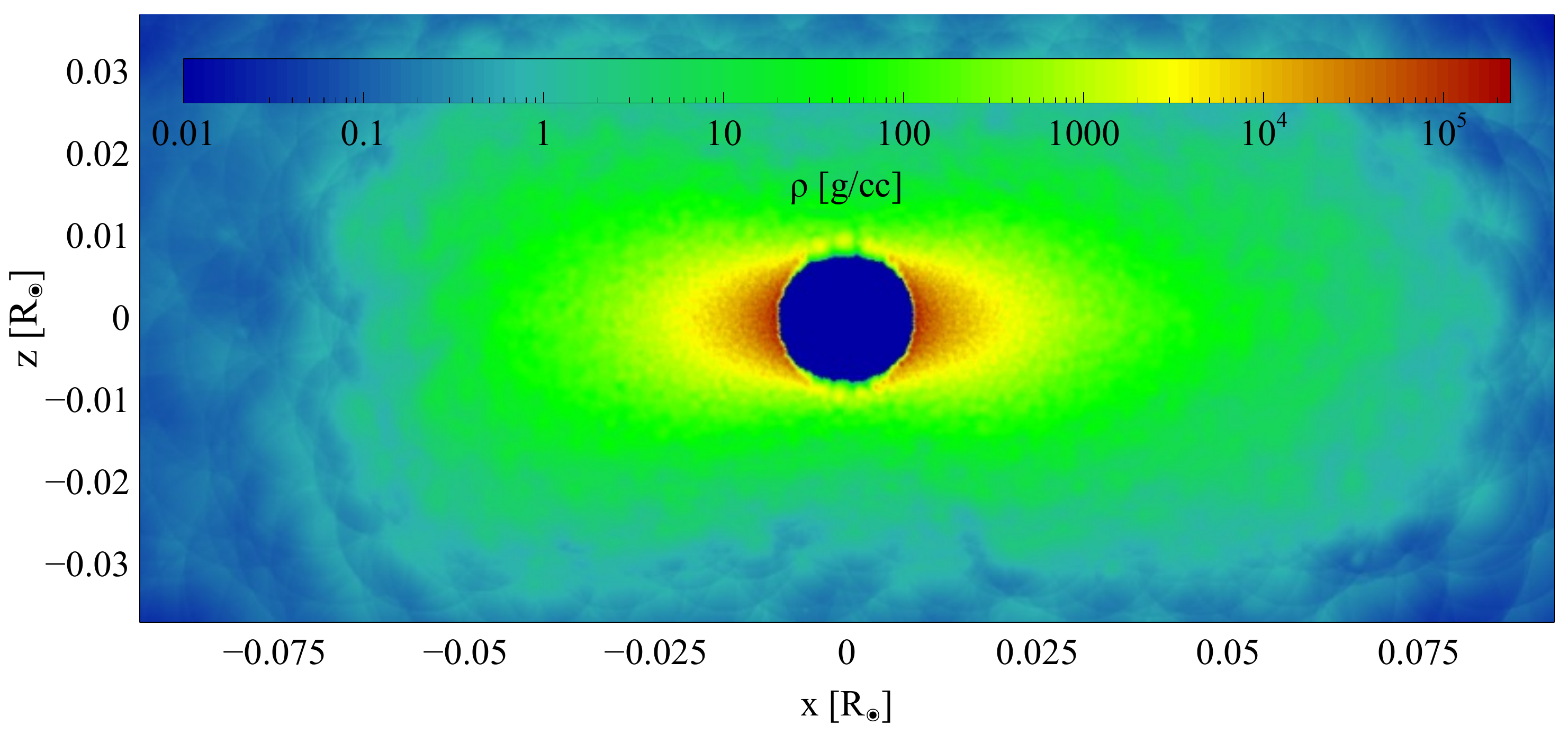}
\caption{A slice through the $x$-$z$ plane of the disk remnant in simulation 5, 0.81\msol$\times2$, with the core removed.}
\label{fig:5disk}
\end{figure}

While the disk is large in spatial extent with an outer radius of $\approx$0.8R\sol, it represents $\approx$35\% of the total system mass, or $\approx$0.57\msol. Moreover, most of the mass of the disk lies much nearer to the core, with the half-mass radius at $\approx$0.02R\sol, as compared to the radius of the core, $\approx$0.008R\sol. This disk is optically thick, and while the timescale for molecular diffusion is of order Myr, radiation at the surface of the disk will drive convection. With convection turning over much of the gas in the disk and acting as a viscosity, angular momentum will be lost from the inner parts of the disk and it will accrete onto the core. We can estimate the timescale for accretion using an $\alpha$-disk prescription for the viscosity (Shakura \& Sunyaev 1973) as 
\be
\tau_{acc}\simeq \alpha^{-1} \left(\frac{r_d}{2h}\right)^2\Omega^{-1},
\ee
where $\alpha$ is a free parameter that relates the viscosity to the speed of sound times the scale height $h$, $r_d$ is the half-mass radius of the disk, and $\Omega$ is the angular speed. As Figure \ref{fig:5disk} shows for an edge-on view of the disk in simulation 5 with the core removed, there remains considerable dense material at high latitudes, with $h/r_d\approx0.1$. For an assumed $\alpha\approx0.1$, we find that the accretion time for these disks is of order $\sim100$ orbital times. This is similar to the timescale derived by van Kerkwijk, Chang, \& Justham (2010) for the results of simulations carried out in Lor\'{e}n-Aguilar \etal (2009).

\begin{figure*}[ht]
\centering
\includegraphics[width=0.85\textwidth]{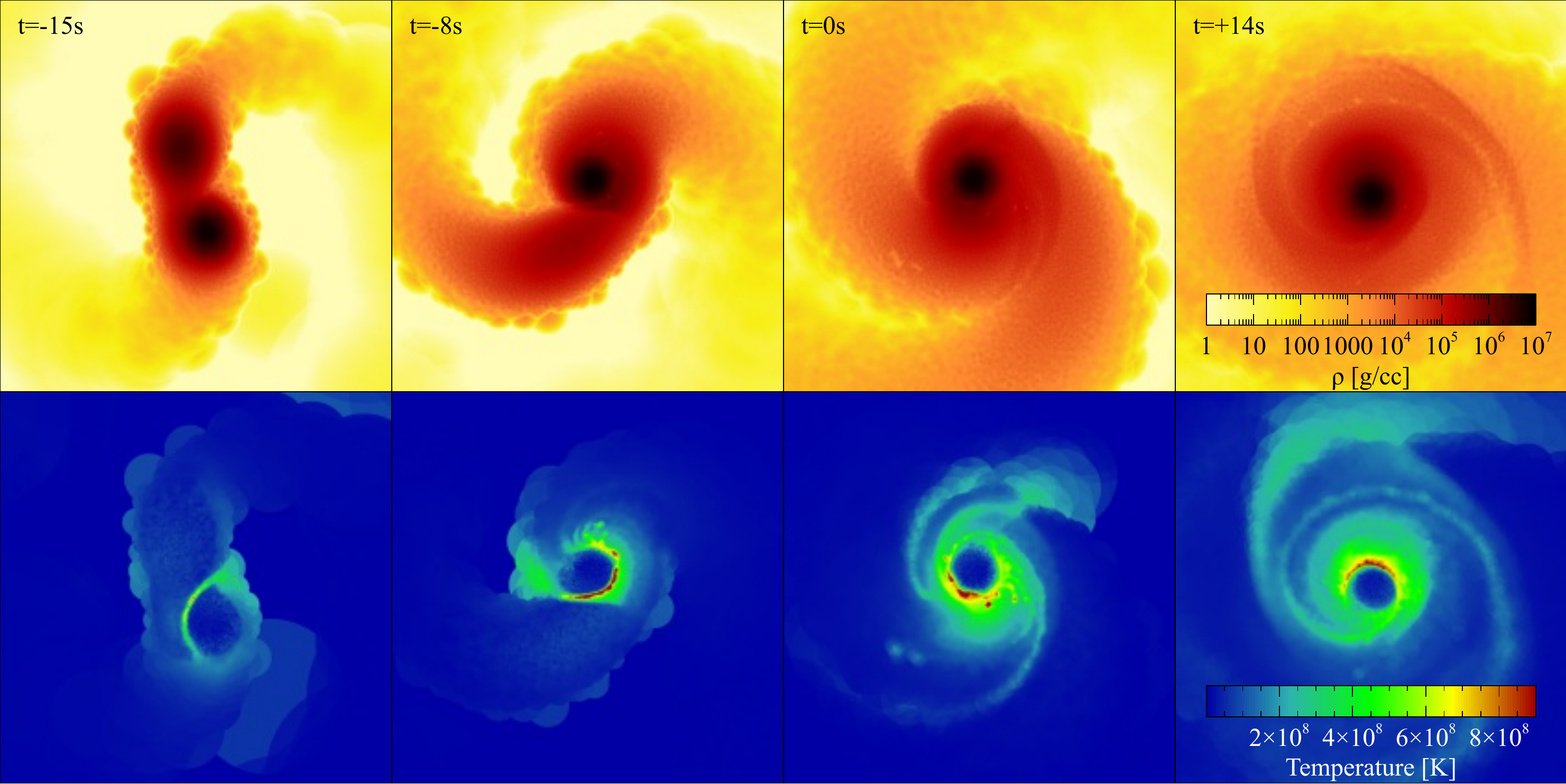}
\caption{Same as figure \ref{fig:0p8x2}, but for simulation 8, 0.96\msol$\times2$.}
\label{fig:0p9x2}
\end{figure*}

Simulation 8 (0.96\msol$\times2$, depicted in Figure \ref{fig:0p9x2}) exhibited very much the same behavior as simulation 5 (0.81\msol$\times2$) with both featuring Mach $>$2 accretion shocks. Here, the larger densities of the constituent stars and faster orbital velocity resulted in greater shock heating, with some of the material reaching temperatures of $\approx$2$\times10^9$ K in a very similar fashion to the simulation studied in Pakmor \etal (2010). However, as evidenced by subsequent snapshots, this shock-heated gas simply expands and cools, joining the flow of material into the disk.

\begin{figure*}[ht]
\centering
\includegraphics[width=0.85\textwidth]{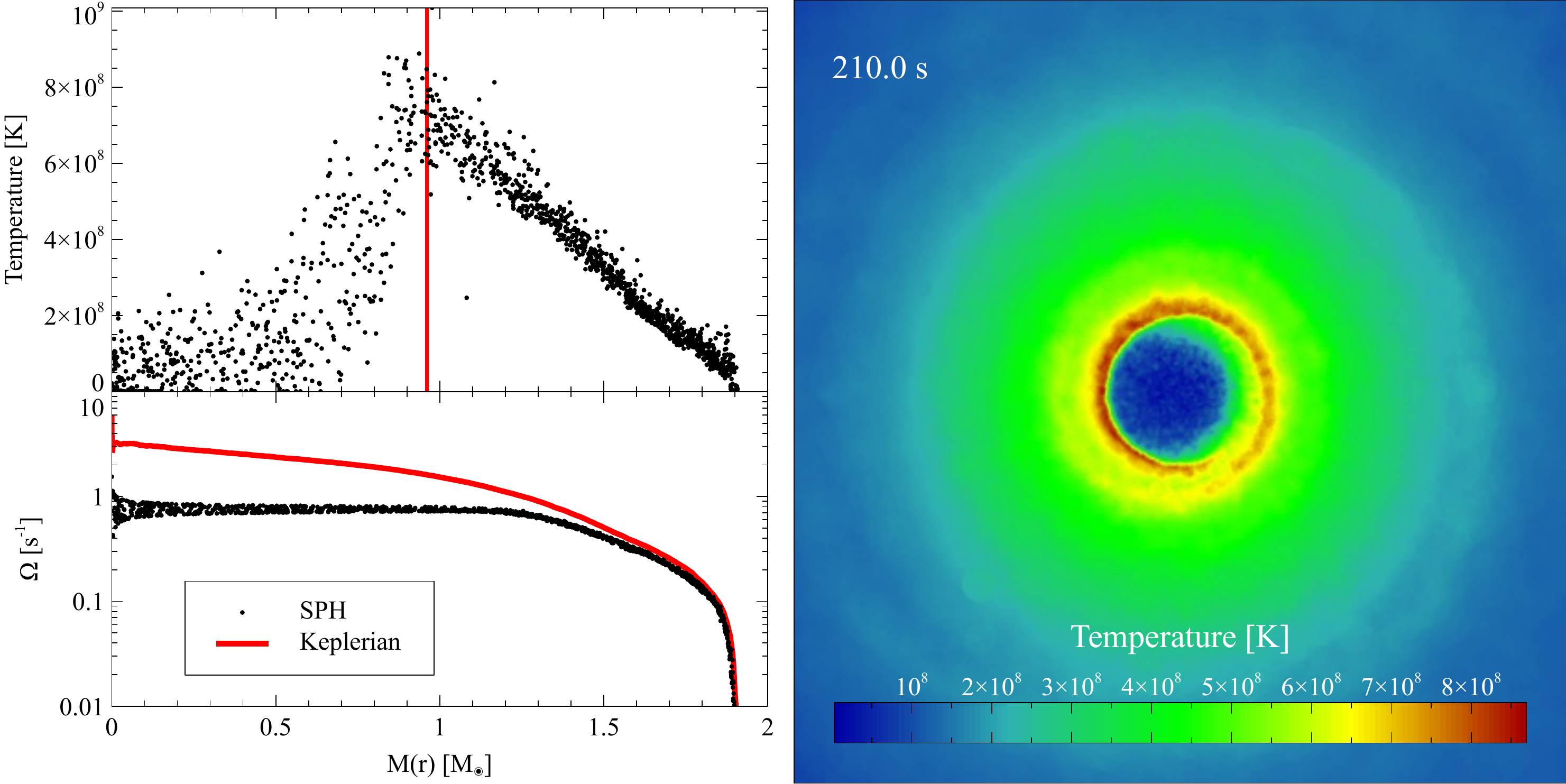}
\caption{Same as figure \ref{fig:5rem}, but for simulation 8, 0.96\msol$\times2$ at 210.0s after the destruction of the secondary.}
\label{fig:8rem}
\end{figure*}

As Figure \ref{fig:8rem} demonstrates, the remnant core is slightly colder and the core-disk interface is slightly hotter than the remnant in simulation 5. This implies that while the shock heating of the disk and interface was stronger, the shock was less able to penetrate through the primary star. In fact, the core of the primary remains essentially unchanged from its original state. 

\begin{figure*}[ht]
\centering
\includegraphics[width=0.85\textwidth]{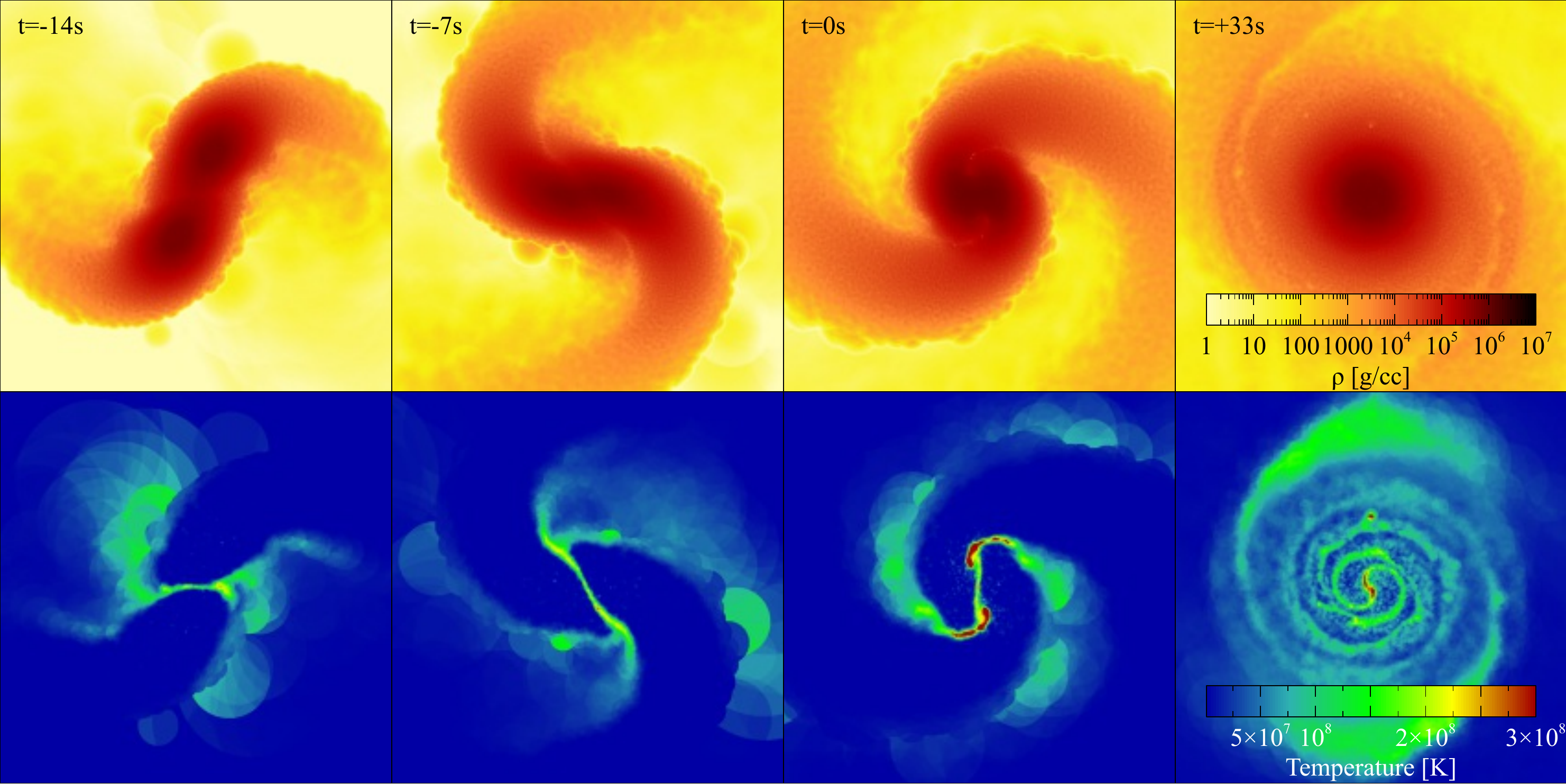}
\caption{Same as figure \ref{fig:0p8x2}, but for simulation 1, 0.64\msol$\times2$.}
\label{fig:0p6x2}
\end{figure*}

For simulation 1, 0.64\msol$\times2$, the common-envelope phase is foreshortened, and unlike the more massive pairs of equal-mass white dwarfs, the cores of both stars merge at the center of mass, as depicted in Figure \ref{fig:0p6x2}. This is most likely due to the stars being less massive and thus, more susceptible to tidal disruption.  

The merged core of the remnant in simulation 1 has a temperature of $\sim2\times10^8$ K and exhibits solid-body rotation out to $\approx$0.84\msol. Roughly 2\% of the initial helium is burned to carbon and oxygen in this simulation, and with $\approx$0.03\msol of helium still remaining in the disk, sub-Chandrasekhar detonation mechanisms that require helium atmospheres may still be viable for this system. 

\begin{figure*}[ht]
\centering
\includegraphics[width=0.85\textwidth]{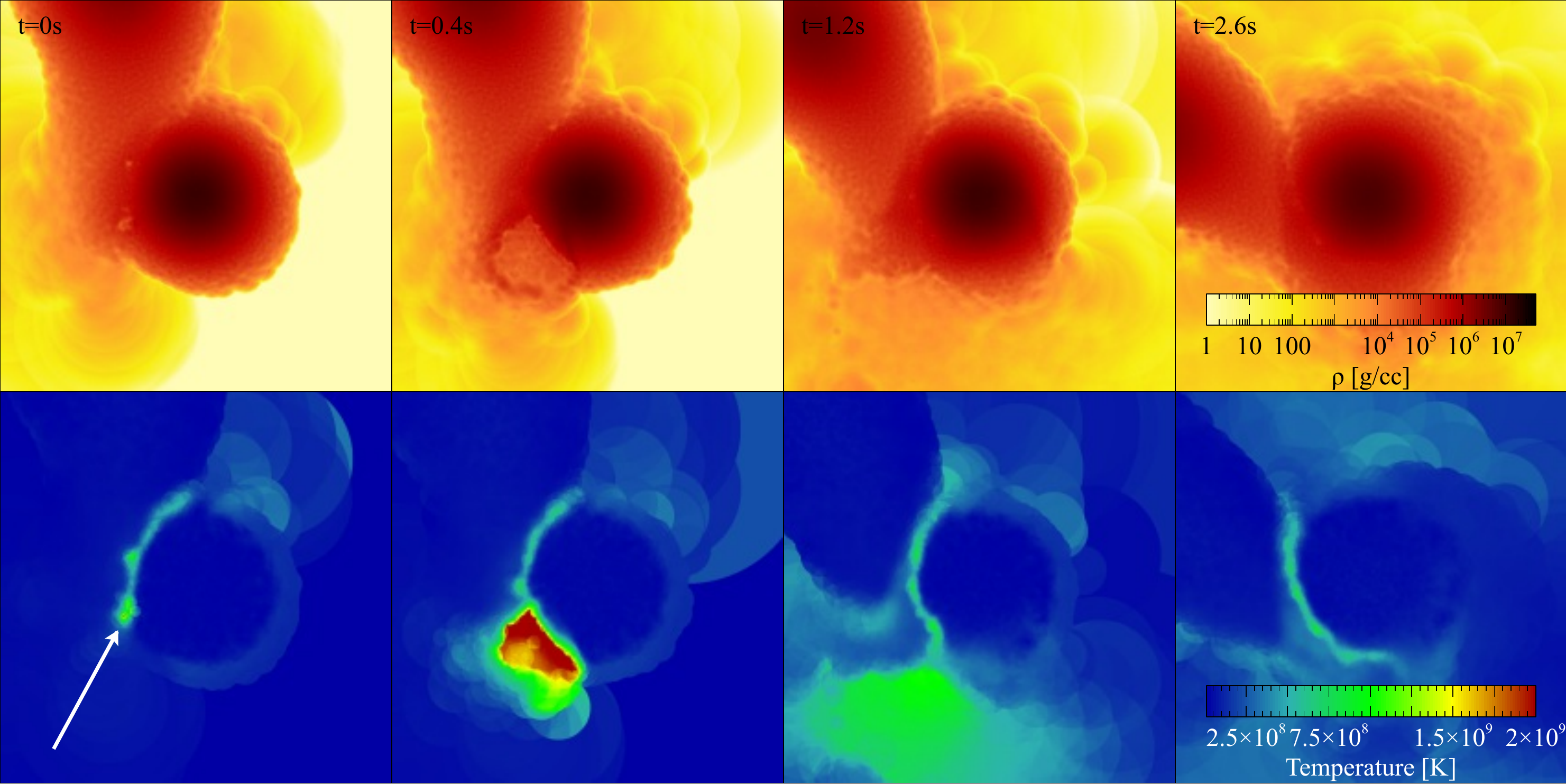}
\caption{Same format as Figure \ref{fig:0p8x2}, but for simulation 9, 0.96\msol + 1.06\msol, with the time coordinate centered on the moment of the helium detonation. The white arrow in the left-most, lower panel indicates the location of the detonation nucleus. A detonation shock can be seen propagating to the right of this location, through the 1.06\msol primary, expanding its outer layers.}
\label{fig:0p9_1p0}
\end{figure*}

For unequal mass simulations, the least massive star is always disrupted entirely, forming an accretion disk around the primary. This is not unlike the equal mass simulations with the more massive constituent stars. However, for mass pairs that involved a 1.06\msol primary, the accretion stream shock on the surface of the primary was sufficiently strong to sustain a helium detonation. This is most clearly evident in simulation 9, 0.96\msol + 1.06\msol, depicted in Figure \ref{fig:0p9_1p0}. 

This detonation burns $\approx$60\% of the helium on the surface of the 1.06\msol primary and propagates a shock through its core, raising its core temperature slightly and expanding its outer layers on the opposite side of the accretion shock. The detonation shock joins the accretion shock in the accretion stream and stalls as the material there is moving at near the sound speed onto the primary. The energy released from the initial detonation is $\sim10^{49}$ erg, however, the detonation was not sufficiently energetic to burn carbon in considerable quantities, or to unbind the primary star. 

Surface helium detonations in white dwarf binary mergers were also observed in grid-based simulations by Guillochon \etal (2010). They found that Kelvin-Helmholtz instabilities were the culprit mechanism for initiating a surface detonation on a 0.9\msol primary, and that the likelihood of a detonation depended very sensitively on the stream geometry and spatial resolution. SPH methods are not well suited for resolving instabilities of this type, and so, given that all of our 1.06\msol primary simulations featured helium detonations, it is likely that for massive primaries, helium detonations are a robustly replicated result. 

This has significant implications for double detonation progenitor models (\eg Nomoto 1982; Woosley \& Weaver 1994; Fink \etal 2010) wherein a helium detonation initiates a shock that ignites a carbon detonation in the core of a sub-Chandrasekhar white dwarf. Fink \etal (2007) found that caustics were responsible for igniting the carbon-oxygen cores in their simulations of 1.0\msol white dwarfs. Our simulations and those of Guillochon \etal (2010) did not result in double detonations, and in our case, we cannot capture caustics at our resolution limit and with the SPH method. Instead, we observed only a single shock propagating through the carbon-oxygen core, launched by the initial helium detonation. However it is likely that for larger primary masses, the single shock we observed would be sufficient to ignite a carbon detonation as the higher central densities of more massive white dwarfs reduce the length-scales required for spontaneous detonation of \carbon[12]/\oxygen[16].

\begin{figure*}[ht]
\centering
\includegraphics[width=0.85\textwidth]{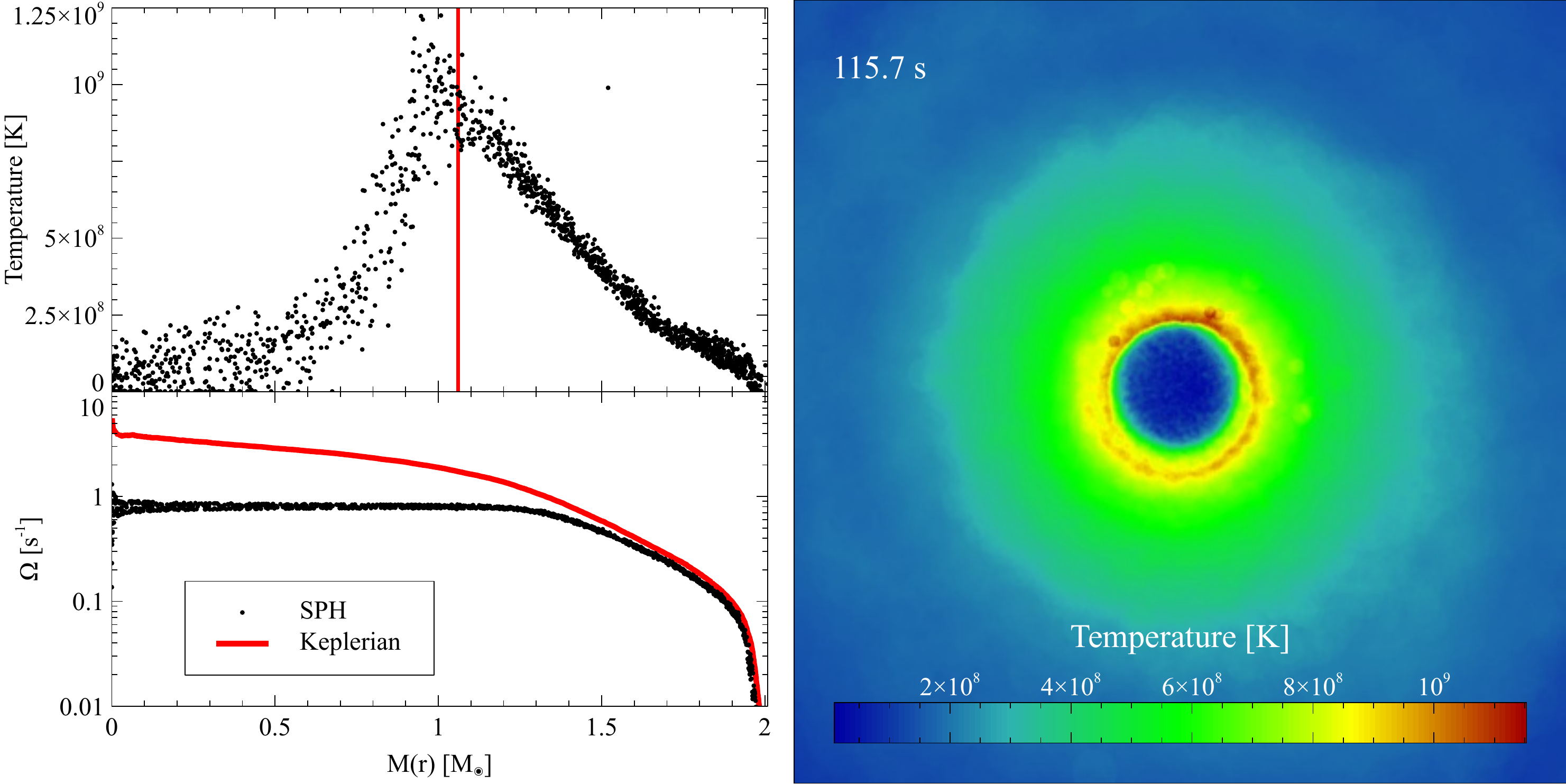}
\caption{Same as figure \ref{fig:5rem}, but for simulation 9, 0.96\msol + 1.06\msol, at 115.7s after the destruction of the secondary.}
\label{fig:9rem}
\end{figure*}

Except for the helium detonation, simulation 9 proceeds in a very similar fashion to the equal mass simulations already discussed. While the 0.96\msol secondary is entirely disrupted into an accretion disk, the core of the 1.06\msol primary is largely unchanged, though slightly heated by the detonation shock. As shown in Figure \ref{fig:9rem}, the core-disk interface is much hotter than previous simulations, reaching a maximum temperature of 1.25$\times10^9$ K. The core exhibits solid-body rotation out to $\approx$1.23\msol.

\begin{figure*}[ht]
\centering
\includegraphics[width=0.85\textwidth]{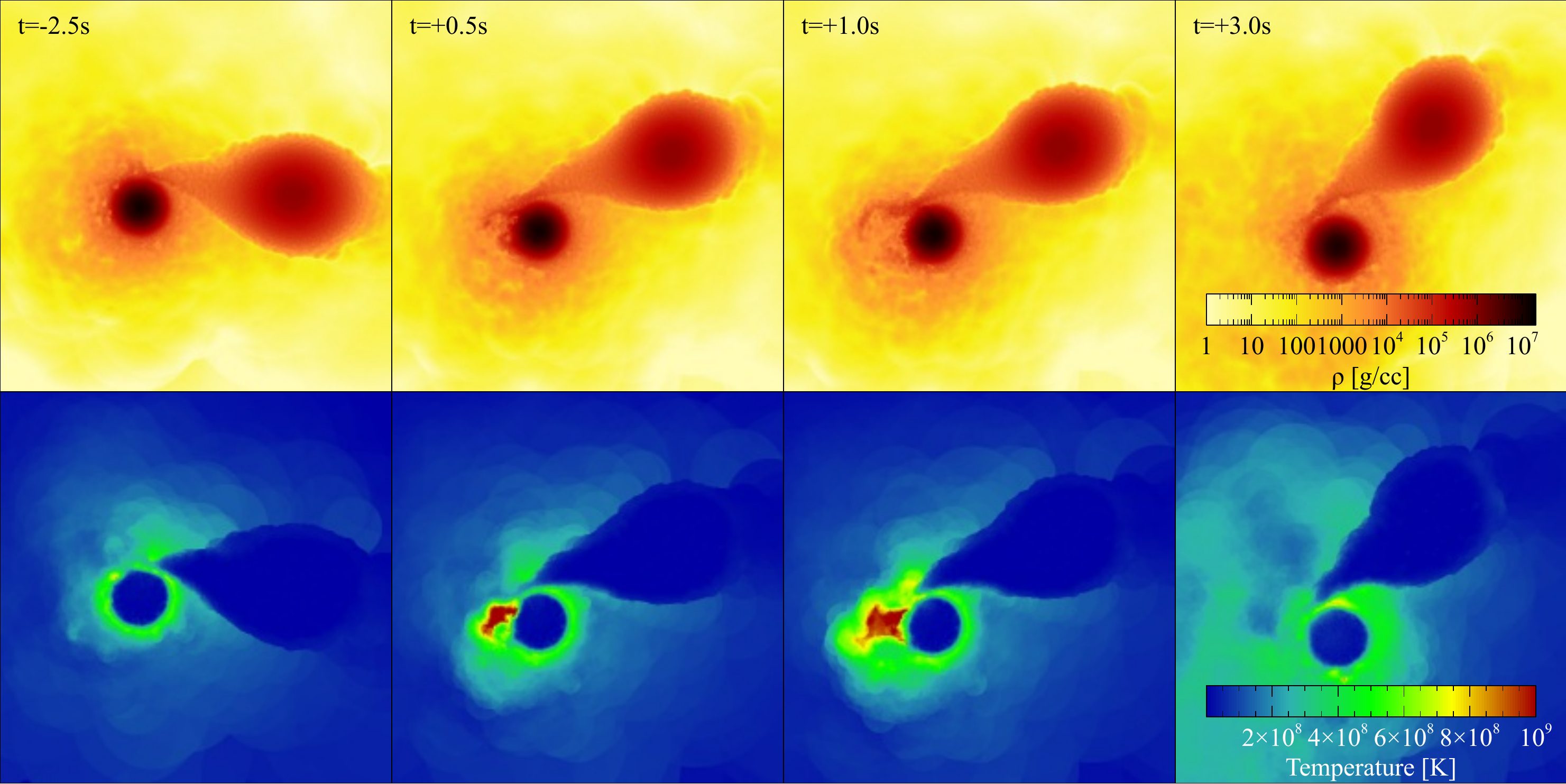}
\caption{Same format as Figure \ref{fig:0p8x2}, but for simulation 4, 0.64\msol + 1.06\msol, with the time coordinate centered on the moment of the helium detonation. A detonation shock propagates to the right, through the 1.0\msol primary, expanding its outer layers.}
\label{fig:0p6_1p0}
\end{figure*}

Simulations 4 and 7, 0.64\msol + 1.06\msol and 0.81\msol + 1.06\msol, both featured these helium detonations as well. The detonation for simulation 4 is illustrated in Figure \ref{fig:0p6_1p0}. As in simulation 9, 0.96\msol + 1.06\msol, the detonation propagated a shock through the primary, expanding its outer layers, but it did not significantly affect the remnant properties or the formation of a disk. The properties of the disk in this simulation and of the disks in each of our simulations are given in Table \ref{table:accretion}.

\begin{table}[ht]
\caption{Simulated binary mass pairs and their disk properties. The half-mass radius of the disk is given as $r_d$. $\Omega$ is the rotational speed of the inner disk. The core mass, $M_{core}$, is quoted with the upper limit representing the mass that the core would grow to if thermal support was removed from the disk. All units are solar unless specified otherwise.}
\centering
\begin{tabular}{c | c c | c c c c}
\hline\hline
\# & $m_1$ & $m_2$ & $M_{disk}$ & $M_{core}$ & $r_d$ & $\Omega$ [s$^{-1}$]\\
\hline
1 & 	0.64 & 	0.64 & 	0.44 & 	0.84-0.99 & 	0.019 &	0.36\\ 
2 & 	0.64 & 	0.81 & 	0.49 &	0.96-1.05 & 	0.023 &	0.38\\
3 & 	0.64 & 	0.96 & 	0.53 & 	1.07-1.16 & 	0.028 &	0.45\\
4 & 	0.64 & 	1.06 & 	0.59	& 	1.11-1.20 &	0.029 &	0.46\\
5 & 	0.81 & 	0.81 & 	0.57 & 	1.05-1.20 & 	0.019 & 	0.47\\ 
6 & 	0.81 & 	0.96 & 	0.67 &	1.10-1.26 &	0.020 &	0.57\\
7 & 	0.81 & 	1.06 & 	0.75 &	1.12-1.30 &	0.019 &	0.69\\
8 & 	0.96 & 	0.96 & 	0.67 & 	1.25-1.44 &	0.016 &	0.67\\ 
9 & 	0.96 & 	1.06 & 	0.79 &	1.23-1.42 &	0.017 &	0.78\\ 
10&	1.06 &	1.06 &	0.87	&	1.25-1.48 &	0.016 &	0.78\\
\hline
\end{tabular}
\label{table:accretion}
\end{table}

In each of the simulation remnants that featured a cold core with a hot disk, a non-negligible fraction of the inner parts of the disks were supported mainly by thermal pressure, as their orbital velocities were sub-keplerian. Moreover, much of the inner portions of the disks had orbital energies below that which would sustain an orbit at the surface of the core. It follows then, that as these disks cool, their inner parts will accrete onto their cores, irrespective of angular momentum transfer, and may do so on rapid timescales. 

The disks of simulations 8-10 retain considerable material ($\apprge$0.2\msol) held aloft only by thermal pressures, suggesting that the cores in those scenarios may grow to super-Chandrasekhar masses from cooling alone, before the disk becomes appreciably thin. If this proves to be the case in future simulations that include radiative cooling, any supernovae that result from these scenarios would appear very much unlike standard SNeIa as they would be enshrouded by thick disk material.

\subsection{Convergence}

\begin{figure}[ht]
\centering
\includegraphics[width=0.45\textwidth]{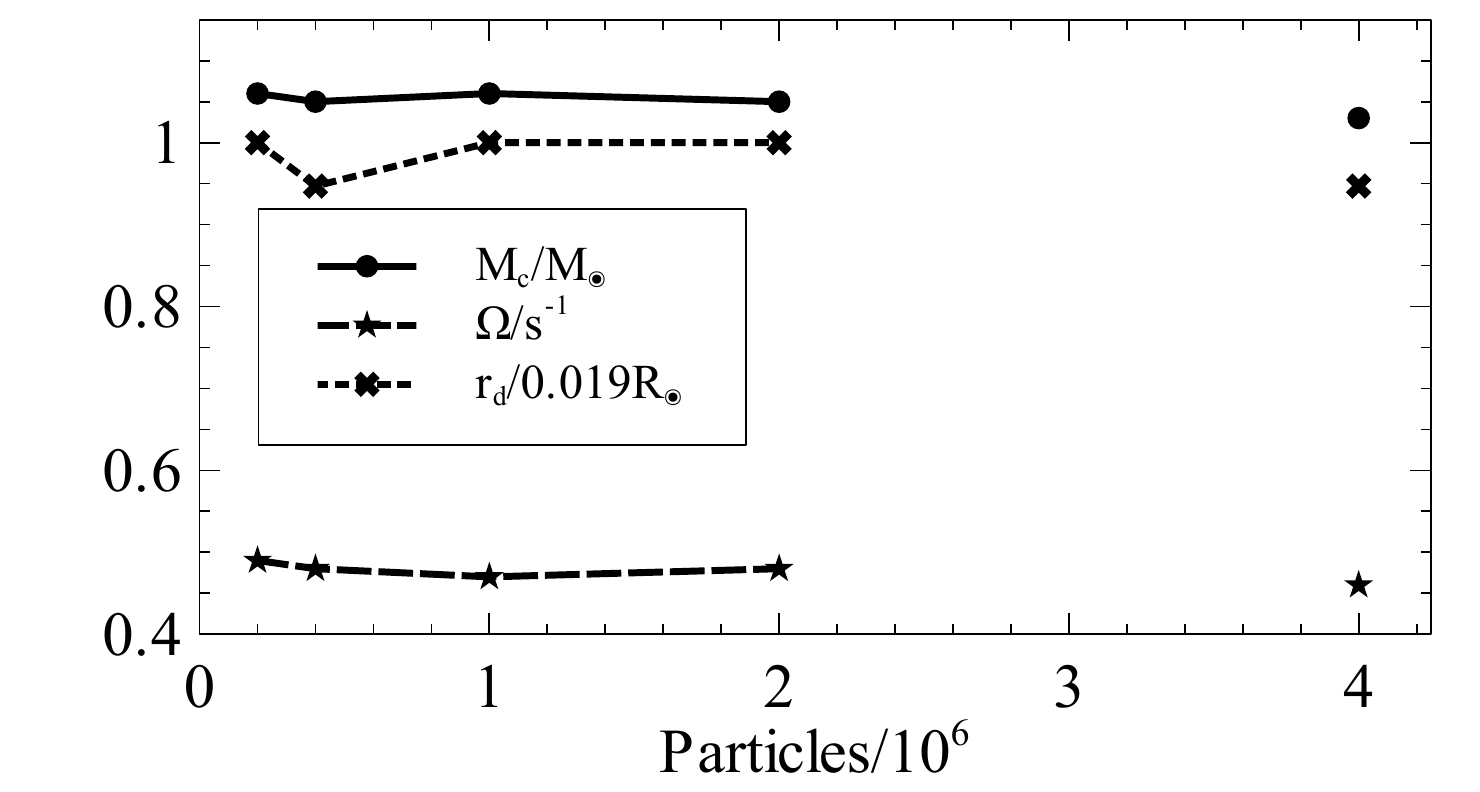}
\caption{A plot of convergence for three variables at various resolutions (total particles) in simulation 5, 0.81\msol$\times2$. The values at $4\times10^6$ particles not connected by lines are those found in the simulation of 0.81\msol + 0.79\msol.}
\label{fig:conv}
\end{figure}

To test convergence of our results, we ran simulation 5, 0.81\msol$\times2$, at various particle resolutions including $10^5$, $2\times10^5$, $5\times10^5$, $10^6$, and $2\times10^6$ particles per star, as this simulation contained equal and representative white dwarf masses, and thus provides the best opportunity to examine resolution effects. In all of our trials, the values for $M_{disk}$, $r_d$, and $\Omega$ differed by about 2\%, as seen in Figure \ref{fig:conv}. From this and the fact that these values did not exhibit a trend in any particular direction, we conclude that the results of our other simulations at $5\times10^5$ particles per star are indeed converged. 

At very high particle counts, such as our simulation with $4\times10^6$ particles total, the asymmetry of the stars stemming naturally from a particle-based method is not sufficient to disrupt one of the stars onto the other. Instead, the system settled into a stable equilibrium with a matter bridge extending between the two stars. Since this configuration is highly unlikely in nature, we reduced the mass of one of the stars by 2\% in the initial condition state to restore an asymmetry. Even in this scenario, the three variables tested for convergence, $M_c$, $r_d$, and $\Omega$ are nearly identical to the previous trials, only reduced slightly due to the reduced total system mass. From this, we conclude that when there is an asymmetry either from numerical noise or enforced by mass constraints, the physics that lead to the final configuration are robust at our resolutions.

\section{Comparison to Previous Studies}

Other groups have simulated white dwarf mergers of various mass pairs and at varying resolutions in the past. Table \ref{table:prev}, while not meant to be exhaustive, summarizes many of the past works on this topic and provides a comparison to the work outlined in this paper.

\begin{table*}[ht]
\caption{A tabular summary of simulation studies of white dwarf mergers.}
\begin{tabular}{c l | c c c c c}
\hline\hline
\# & Reference & EOS & Network & Masses [\msol] & Particles\\
\hline
1 & Benz \etal 1990			& Chandra$^\alpha$	& --- 				& 0.9+1.2 		& $3\times10^3$\\
\hline
2 & Rasio \& Shapiro 1994	& Polytrope 			& --- 				& --- 			& $4\times10^4$\\
\hline
3 & Segretain \& Chabrier 1997	& Chandra			& --- 				& 0.6+0.9 		& $5.83\times10^4$\\
\hline
4 & Guerrero \etal 2004		& 3 press.$^\beta$		& 14 isotope			& 0.6+0.8 0.6+1.0 			& $4\times10^4$\\
   &					&				&				& 0.8+1.0\\
\hline
5 & Yoon \etal 2007			& Helmholtz$^\gamma$	& QSE+$\alpha$-chain 	& 0.6+0.9		& $2\times10^5$\\
\hline
6 & Lor\'{e}n-Aguilar \etal 2009	& 3 press.			& 14 isotope			& 0.3+0.5 0.4+0.8	& $4\times10^5$\\
   &						&				&				& 0.6+0.6 0.6+0.8\\
   &						&				&				& 0.6+1.2\\
   \hline
7 & Pakmor \etal 2010, 2011	& Helmholtz			& 13 isotope			& 0.89+0.89 0.7+0.9 & $10^6$\\
   &						&				&				& 0.76+0.9 0.81+0.9\\
   &						&				&				& 0.89+0.9\\
   \hline
8 & Dan \etal 2011			& Helmholtz 			& 7 isotope 			& many$^\delta$	& $4\times10^4-2\times10^5$\\
\hline
9 & This Work				& Helmholtz 			& 13 isotope			& see table \ref{table:grid} & $10^6$\\
\hline
\end{tabular}
{\raggedright
\newline$^\alpha$ Chandrasekhar 1939
\newline $^\beta$ Ideal gas + Fermi Pressure + Radiation
\newline $^\gamma$ Timmes \& Arnett 1999
\newline $^\delta$ 0.2+0.8 0.3+1.1 0.5+1.2 0.3+0.6 0.6+0.9 0.2+0.3 0.3+0.4 0.9+1.2 0.45+0.9 0.45+0.67}
\label{table:prev}
\end{table*}	

Groups 1-3 completed pioneering work, modeling white dwarf mergers in smooth-particle hydrodynamics codes. Their particle resolutions were quite low compared to what is possible with current computing resources, and their equations of state were mainly analytical approximations for degenerate matter. They also did not include any nuclear reaction networks in their calculations. Rasio \& Shapiro (1994) consider the hydrodynamics of binary
coalescence between two polytropic stars using a rotating-frame
potential.  Accounting for tidal distortions on the separation
distance for Roche overflow, and thus capturing more realistic
accretion rates, they find merger times on the order of minutes,
rather than tens of seconds as other groups had found.  Rasio \&
Shapiro (1994) did not explicitly list the masses of their binary
configurations since the hydrodynamical response using a polytropic
EOS should depend only on mass ratio.

With improved computing power, merger calculations began to incorporate more accurate equations of state and to include nuclear reaction networks such as those used by groups 4-6. Resolutions were also moderately improved, and more mass combinations were explored such as those of Lor\'{e}n-Aguilar \etal (2009). And while Lor\'{e}n-Aguilar \etal (2009) simulated what was, up to then, the largest sample of mass combinations, each of the groups numbered 4-6 used approximate initial conditions that started the stars much closer than in Rasio \& Shapiro (1994), which had the result of overestimating the accretion stream and underestimated the breakup time of the secondary star. Lor\'{e}n-Aguilar \etal (2009) also did not attempt mass combinations where both stars had $M>0.6\msol$. Moreover, in their conclusions, Yoon \etal (2007) singled out three important factors that would be important for subsequent calculations: 1) the effects of tidal forces on the heating of the stars and disk, 2) the presence of hydrogen/helium atmospheres, and 3) a larger range of mass combinations that can lead to varying final remnant configurations.  All of these are taken account in the present work.

More recently, Pakmor \etal (2010,2011) simulated a wide variety of almost exclusively high-mass pairs with the aim of studying the outcomes of early detonations in white dwarf mergers. However, as stated previously, their initial conditions were approximate and prone to overestimating the accretion rate of the disk material. 

The work of Rasio \& Shapiro (1994) and of Dan \etal (2011) motivated our choice of initial conditions. This group also studied a wide range of mass combinations, but as their primary focus was on Roche overflow and accretion rates, their resolutions were comparably low and their nuclear network was minimal. In the work outlined in this paper, we have attempted to combine the best features of all of these previous works including high resolutions, a wide range of mass combinations (and high-mass pairs), a robust equation of state and nuclear network, and accurate initial conditions that properly model accretion rates. We have also included helium atmospheres in order to gauge how robust helium detonations are on the surfaces of the primary stars, and to what extent those detonations affect the final merger remnants. 

\section{Discussion}

Double degenerate progenitor scenarios are drawing new interest among the supernova community. They have the potential to explain some of the confounding mysteries that remain about SNeIa, and to enhance their usefulness as cosmological probes. Here, we have conducted a large survey of white dwarf binary merger models in order to begin to understand how these systems evolve, and to piece together some of the observational signatures of double degenerate SNeIa. 

In each of our simulations, the merger remnant consisted of a cold, degenerate core surrounded by a hot accretion disk, with $h/r_d\approx0.1$. The only exception to this remnant configuration was the simulation of two 0.64\msol white dwarfs, where the cores of both stars merged at the center of mass, heating the remnant core considerably and lifting most of its degeneracy. Since it is unlikely that two identical mass white dwarfs would form in a binary in nature, this scenario might seem of trivial importance, but the merging of the cores was less the result of their masses being identical than it was of the stars being highly susceptible to tidal disruption. Our other equal mass simulations with more massive constituent stars featured merger scenarios wherein one of the stars was completely disrupted into an accretion disk around the core of the other, provided there was a non-trivial asymmetry between the two. Therefore, it is likely that merger scenarios involving slightly unequal but low-mass white dwarfs might also exhibit core merging, though more simulations are certainly needed to confirm this conjecture. 

Evolving these meta-stable remnants further will require the implementation of new physics beyond hydrodynamics and nuclear burning. Since the disks are optically thick, radiative losses from the surface are the dominant evolutionary mechanism, and efforts are underway to incorporate this process. As the surface layers lose energy through radiation, the material will sink to lower latitudes, driving convective currents that will increase the disk viscosity. We expect this viscosity will mimic an $\alpha$-disk prescription, and thus, the accretion times will be of order 100 orbital times. However, the three most massive combinations simulated have disk configurations that suggest the cores might grow to become super-Chandrasekhar masses on shorter timescales, after cooling has removed some of the thermal support for the inner portions of the disks.

While none of our simulations exhibited prompt carbon detonations, all of the simulations that included a 1.06\msol primary did exhibit prompt helium detonations on the surface of the primary. These helium detonations were not sufficiently energetic to significantly burn much of the carbon or to unbind either of the constituent stars, but the detonation shocks they produced did alter the structure of the primary before complete merger. Moreover, at $\sim10^{49}$ ergs, the energy these detonations released is likely sufficient to be observable by many of the upcoming transient surveys, such as LSST. 

Whether white dwarf mergers produce SNeIa is still an open question. Our simulations and others' represent only the beginnings of our exploration of this progenitor mechanism. Preliminary results look very promising, and it is doubtless that the viability of mergers as SNeIa progenitors will be established in the near term. Meanwhile, it remains an exciting time to be exploring these dynamical scenarios that continue to surprise us.


\section*{Acknowledgments}

This work was supported by the National Science Foundation under grant AST 08-06720, by the National Aeronautics and Space Administration under NESSF grant PVS0401, and by a grant from the Arizona State University chapter of the GPSA. All simulations were conducted at the Advanced Computing Center at Arizona State University. We also thank Peter Hoeflich, Philip Chang, Marten van Kerkwijk, Dan Kasen, Stephan Rosswog, James Guillochon, Enrico Ramirez-Ruiz, Ken Shen, Andy Nonaka, and Ann Almgren for insightful discussions during the creation of this manuscript.

\end{document}